# Büttiker Probe Based Modeling of TDDB: Application to Dielectric Breakdown in MTJs and MOS Devices

Ahmed Kamal Reza, Mohammad Khaled Hassan and Kaushik Roy

*Abstract*— **Dielectric layers are gradually being downscaled in different electronic devices like MOSFETs and Magnetic Tunnel Junctions (MTJ) with shrinking device sizes. As a result, time dependent dielectric breakdown (TDDB) has become a major issue in such devices. In this paper we propose a generalized way of modeling the stress induced leakage current (SILC) and post breakdown current (PBC) due to time dependent wear-out of the dielectric layer. We model the traps formed in dielectric layer using Büttiker probe [1] and incorporate the Büttiker probe self-energies in standard self-consistent Non-Equilibrium Green's Function (NEGF) formalism [2] in order to determine SILC and PBC. In addition, we have shown the impact of break down in the dielectric layer on the spin current and spin filtering characteristics of an MTJ. The proposed model is generic in nature. It can be extended from MTJs and conventional CMOS technology to any other devices with any type of single and multiple layers of dielectric material(s).**

*Index Terms*— **Magnetic tunnel junction (MTJ), Time dependent Dielectric Breakdown (TDDB), Spin Current Degradation, Non-equilibrium Green's function (NEGF), Büttiker probe, Stress Induced Leakage Current (SILC), Soft Breakdown, Hard Breakdown.**

## I. INTRODUCTION

E XTENSIVE research is currently being done on spin based memory, logic and neuromorphic computing devices and circuits. Today, MTJs are becoming an indispensable part in almost all spin-based systems. In order to achieve better write current, researchers are now fabricating MTJs with ~1nm thick MgO layer in between the fixed and the free ferromagnetic layers [3]. As a result, the voltage applied across these thin oxide layers can generate a tremendous amount of stress electric field (on the order of $10^8$-$10^9$ V/m). Gradually traps start forming in the dielectric layer that eventually lead to the formation of percolation paths. When electron is captured by a trap, the spin orientation of that electron is randomized [4] [5]. As a result, the spin filtering efficiency of the MTJ starts degrading over time, eventually causing functional failures.

Post-breakdown current-voltage (I-V) characteristic primarily depends on the type of dielectric breakdown. If the post-breakdown I-V characteristic is vastly different from its no-breakdown counterpart and follows Ohm's law [6], the dielectric is said to have experienced a hard breakdown. On the other hand, soft dielectric breakdown is characterized by a power law dependence between post-breakdown current and the corresponding voltage [6]. But in this case the increase in post breakdown current at low voltage is smaller compared to corresponding hard breakdown current [6]. Over the past decade a lot of research [6-13] focused on the breakdown characteristics/models of dielectric layers, especially for MOS devices. These mainly analytical in nature and can predict the breakdown behavior such as, post break down I-V characteristics and the time to failure. However, the analytical models are not sufficient to predict the spin current degradation through MTJ due to dielectric breakdown. This is because the ferromagnetic layers can be either in parallel or anti parallel states (fig. 1) and the tunneling ferromagnetic resistance varies with the corresponding magnetization orientation (fig. 2). Moreover, the spin tunneling current varies a lot depending on the angle of magnetization of fixed and free layer in an MTJ (fig. 2). In addition, different ferromagnetic layers can have different exchange coupling energy and different band structures that play a significant role in determining the post breakdown MTJ characteristics. Therefore, it is evident that for reliable spin dependent current simulation, we need to explicitly consider both the up spin and the down spin band diagrams and the density of states of both the ferromagnetic magnetic contacts and the channel dielectric.

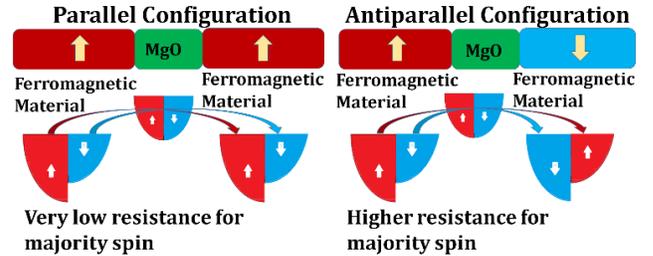

Fig 1: Schematic band diagram of CoFeB-MgO-CoFeB MTJ for both parallel and anti-parallel configurations. Red and blue color represents up and down spin respectively.

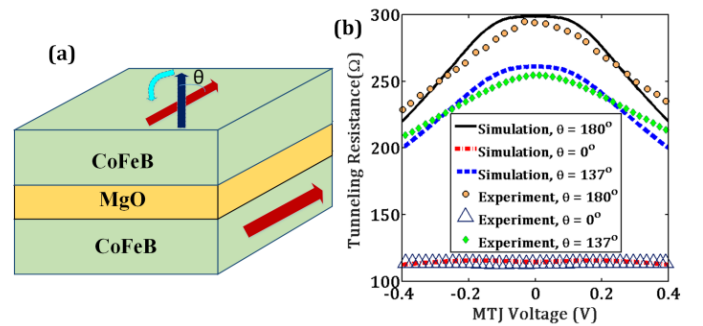

Fig 2 (a). Schematic diagram showing the angle between the magnetization of fixed and free layer in CoFeB/MgO/CoFeB MTJ. The angle is defined as θ. (b) Simulation data with our NEGF framework (without any defect or breakdown) and experimental data [3] on variation of MTJ resistance with changing θ. Three values of θ (θ = 0°, 137° and 180°) are considered here. (Root mean square value of error is less than 4%)

A.K. Reza, M.K. Hassan and K. Roy are with the School of Electrical and Computer Engineering, Purdue University, West Lafayette, IN 47907-1285, USA (email: areza@purdue.edu).

For modeling the post breakdown gate current-gate voltage characteristics, various theories have been proposed in the literature. They include variable range hopping (VRH) [14], inelastic macroscopic quantum tunneling [15], and quantum point contact (QPC) model [16]. In VRH theory, post soft breakdown current is modelled by an empirical equation $I_g = aV_g^b$ (because soft breakdown $I_g$-$V_g$ characteristic follows the power law), where, a and b are two empirical parameters depending on the type and the thickness of gate dielectric material. Soft breakdown spin current can also be modeled by such empirical equation. However, every time the material or the dimension of contact ferromagnet or channel dielectric are changed, one has to do a series of experiments to determine a and b. Even for fixed material and device dimensions, the relative magnetization orientation of the fixed and the free layer plays a large role in MTJ I-V characteristics. Inelastic macroscopic quantum tunneling method defines the tunneling resistance as $R_{ti}^{-1} = \frac{4\pi e^2}{h}|T^i|\zeta_i\zeta_0$ [15]. Here $T^i$ is the tunneling matrix and $\zeta_i$ and $\zeta_0$ are density of states in the electrodes. The gate tunneling current is defined as [15],

$$I_g(V_g) = \frac{2\pi e}{h}(\prod_{i=1}^{N}\frac{R_Q}{4\pi^2 R_{ti}})S^2\sum_{k=0}^{N-1}c_k^N(k_BT)^{2k}(eV_g)^{2(N-k)-1}$$

$R_Q$ is the quantum resistance, N is the number of tunnel junctions and T is the temperature. This method cannot calculate the change MTJ current change due to dielectric breakdown because it does not consider spin up and down states separately. Also in this model the breakdown current variation with temperature is modeled using a separate empirical relation $I_g(V_g) = aV_g^9 + bV_g^7T^2 + cV_g^5T^4 + d\,V_g^3T^6 + eV_g\,T^8$ where a, b, c, d and e are fitting parameters with T being the temperature. In our proposed NEGF based modeling, the effect of temperature is already built in the Fermi-Dirac distribution function (defined as $f = \frac{1}{1+exp^{E-E_F/kT}}$, $E_F$ being the Fermi level). Hence, there is no need to model temperature variation separately. In QPC model, the post breakdown current is modeled as [16],

$$I_g(V) = \frac{2e^2}{h}(V-V_0) + \frac{1}{\alpha}\ln[\frac{1+exp^{\alpha(\phi(T)-\beta e(V-V_0))}}{1+exp^{\alpha(\phi(T)+(1-\beta)e(V-V_0))}}]$$

where $\alpha$ is fitting parameter depending on the shape of the energy barrier, V0 is the voltage drop at the two electrodes, β is the fraction of voltage drop cross oxide and, $\phi$, the barrier height is given by $E_0$-$E_F$, where $E_0$ is the bottom of first sub-band. The tunneling current is modeled using the analytical equation. Note, however, that this is not sufficient to model the spin based tunneling current since it does not consider the spin dependent density of states. Another drawback of QPC model is that it projects the temperature dependence of post breakdown current using analytical equation: $\phi(T) = \phi_0 + \xi T$, where $\xi$ is an empirical fitting parameter. In our model, as stated earlier, we do not need to model temperature dependence using fitting parameters.

In this paper we propose a unified model for predicting the I-V characteristics of the dielectric layer due to formation of traps at different positions and at different energy levels. In the next section we discuss our simulation framework and show how Büttiker probe [1] can be used to model the traps. We have shown that our model can predict the behavior of the dielectric layer after both soft and hard breakdowns. We have verified our model with the experimental results of post breakdown I-V characteristics of both conventional MOSFETs (with both single $SiO_2$ layer and HfSiON/$SiO_2$ multilayer as gate oxide) and MTJ. We also used NEGF formalism to calculate the spin current for both parallel and antiparallel states in order to determine the degradation of the tunneling magnetoresistance (TMR) due to the TDDB effect in MgO layer in MTJs. Finally, we have used the standard 3-D cell based percolation model [12] to predict the MTJ lifetime.

## II. PROPOSED SIMULATION FRAMWORK USING BÜTTIKER PROBE

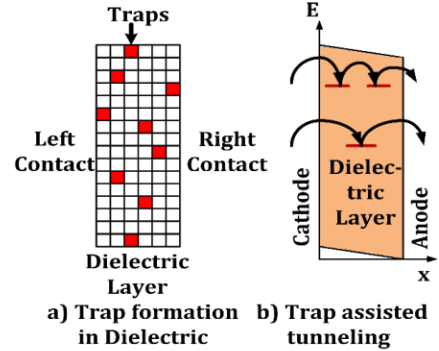

Fig.3.: Schematic diagram of trap position (a) Physical positions of trap (b) Energy level positions of traps.

In this section we first discuss some of the major attributes of trap assisted tunneling. The capture cross section of a trap in a particular dielectric depends on its positon in the energy level [13]. A mid-bandgap trap has a relatively bigger capture cross-section than the ones near the conduction band edge or in the conduction band [17]. Researchers usually measure it experimentally and we will use those experimentally measured values in our analyses. Another important thing to consider is the physical position of the trap. The physical position of a trap determines the distance that a carrier has to tunnel before being captured by the trap. Büttiker probe based NEGF model can be used to determine whether a trap is surface trap or bulk trap by matching the post breakdown I-V characteristic.

In figure 3 we have shown a schematic diagram of both the physical position and energy level of traps for an example dielectric. A major characteristic of trap assisted tunneling is the dephasing of the electron or hole when captured by a trap. Therefore, the tunneling probability and the dephasing of carrier need to be modeled simultaneously.

### A. Modeling traps with Büttiker probe

Büttiker probes are virtual probes that absorb the carriers, dephase them, and inject them back into the channel. Trap assisted tunneling can be modelled using these Büttiker probes by placing one probe at each of the trap location. Similar to the trap assisted tunneling, the carriers can tunnel from one contact to the probe and then from probe to another contact. In addition, the carriers can also tunnel from one probe to another. A Büttiker probe's self-energy matrix can be written according to corresponding trap's physical and energy level position and trap's capture cross section. Thus a Büttiker probe can replicate the functionality of a trap.

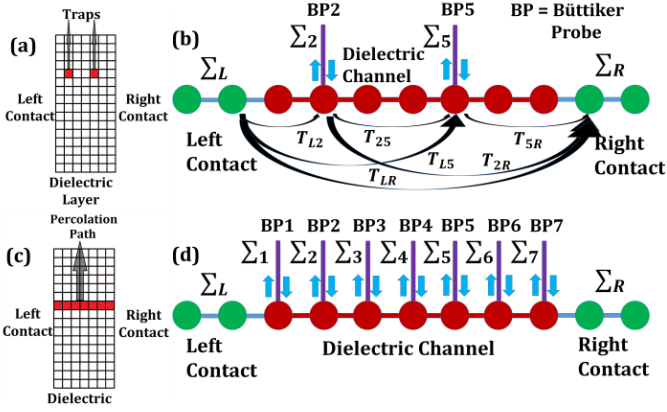

Fig. 4. Concept of Büttiker probe (BP). (a) Consider traps at grid position 2 and 5 (b) Modeling of trap assisted tunneling. Assuming traps are formed at grid position 2 and 5, we have attached Büttiker probes to those positions. All possible tunneling paths are shown with corresponding tunneling transmission probability (c) Formation of percolation path inside the dielectric (d) Modeling of percolation path. A virtual contact is attached to grid points where trap is generated. These additional contacts are treated with a self-energy, like the real contacts.

Fig. 4 explains the concept of Büttiker probes. In addition to Büttiker probes, there are two more contacts (left and right contact) attached to the dielectric layer. In case of MTJ, the left and right contacts are the two ferromagnetic layers, i.e., the fixed layer and the free layer [18]. For MOSFETs, these two contacts are the top metal or polysilicon layer and the bottom channel or the substrate layer. For both cases, as the channel is an insulator, the electrons mainly tunnel through the dielectric layer from one contact to the other. Initially, we will describe the NEGF modeling of direct tunneling current. In order to do that we assumed that there are no pre-existing traps in the dielectric layer. After application of a voltage across the dielectric layer, the Fermi functions at the left and the right contacts are $f_L(E)$ and $f_R(E)$, respectively. The current density can be written as [2] [19].

$$J = \frac{1}{q} \int G(E) \left(f_R - f_L\right) dE \tag{1}$$

Here G(E) is the conductance and q is the electron charge. G(E) is defined in [2] [19] as $G(E) = \frac{q^2}{h} T(E)$, where T(E) is the transmission probability. The transmission probability depends on the retarded Green's function of the system and the self-energy matrices of the contacts. The retarded Green's function can be written as follows [2] [19]

$$\boldsymbol{G^R} = [\boldsymbol{EI} - \boldsymbol{H} - \boldsymbol{\Sigma_L} - \boldsymbol{\Sigma_R}]^{-1} \tag{2}$$

where, $\boldsymbol{\Sigma_L}$ and $\boldsymbol{\Sigma_R}$ are the self-energy matrices of the left and the right contacts, respectively.

Therefore, the transmission probability between the left and the right contacts can be represented as [19]

$$T_{LR} = Trace \left[\boldsymbol{\Gamma_L G^R \Gamma_R G^A}\right] \tag{3}$$

Where, $\boldsymbol{G^A}$ is the complex conjugate matrix of $\boldsymbol{G^R}$. Here $\boldsymbol{\Gamma_L}$ and $\boldsymbol{\Gamma_R}$ are two quantities defined as [19] $\boldsymbol{\Gamma_L} = i[\boldsymbol{\Sigma_L} - \boldsymbol{\Sigma_L^\dagger}]$ and $\boldsymbol{\Gamma_R} = i[\boldsymbol{\Sigma_R} - \boldsymbol{\Sigma_R^\dagger}]$. $\boldsymbol{\Sigma_L^\dagger}$ and $\boldsymbol{\Sigma_R^\dagger}$ are the complex conjugate matrices of $\boldsymbol{\Sigma_L}$ and $\boldsymbol{\Sigma_R}$, respectively. $\Gamma$ is a matrix that physically represents how easily carriers get in or get out of a contact [19].

The concept of Büttiker probe is embedded into NEGF formalism to model the trap assisted tunneling. Let us assume that traps are formed at grid position 2 and 5 (fig. 4(a)). For modeling these two traps, we need to attach two Büttiker probes at those two grid points as shown in figure 4(b). These probes absorb carriers and inject them back to the channel. As a result, there is no net current through these probes i.e., the current conservation law is followed.

Note that the electrons have multiple paths for going from the left contact to the right contact. Electrons can be captured by the trap at grid point 2, then recaptured by the trap at grid point 5 and finally escape through the right contact (fig 4(b)). Also electrons can tunnel from left contact to either traps at grid position 2 or grid position 5 and then tunnel to the right contact (fig 4(b)). Moreover, they can directly tunnel from left contact to right contact. Let us consider the first trap assisted tunneling path. The corresponding tunneling transmission probabilities are $T_{L2}$, $T_{25}$ and $T_{5R}$ (fig 4(b)). These tunneling paths are in series connection with one another and the tunneling transmission probability is proportional to conductance. Therefore, the total tunneling transmission probability of this path, $T_{L25R}$ can be written as

$$\frac{1}{T_{L25R}} = \frac{1}{T_{L2}} + \frac{1}{T_{25}} + \frac{1}{T_{5R}} \tag{4}$$

Where $T_{L2} = Trace \left[\boldsymbol{\Gamma_L G^R \Gamma_2 G^A}\right]$, $T_{25} = Trace \left[\boldsymbol{\Gamma_2 G^R \Gamma_5 G^A}\right]$ and $T_{5R} = Trace \left[\boldsymbol{\Gamma_5 G^R \Gamma_R G^A}\right]$ [2]. If $\boldsymbol{\Sigma_2}$ and $\boldsymbol{\Sigma_5}$ are the self-energy matrices of the Büttiker probes attached at grid positions 2 and 5, then $\boldsymbol{\Gamma_2} = i[\boldsymbol{\Sigma_2} - \boldsymbol{\Sigma_2^\dagger}]$ and $\boldsymbol{\Gamma_5} = i[\boldsymbol{\Sigma_5} - \boldsymbol{\Sigma_5^\dagger}]$. In simple effective mass simulation the we have calculated the self-energy matrix of Büttiker probe as follows

$$\boldsymbol{\Sigma_{Büttiker}} =$$
$$\begin{bmatrix} \left(-t_{de} - qV_{eff}^\alpha\right)\exp(ika) & 0 \\ 0 & \left(-t_{de} - qV_{eff}^\alpha\right)\exp(ika) \end{bmatrix} \tag{5}$$

The corresponding $\boldsymbol{\Gamma_{Büttiker}}$ matrix can be defied as $\boldsymbol{\Gamma_{Büttiker}} = i[\boldsymbol{\Sigma_{Büttiker}} - \boldsymbol{\Sigma_{Büttiker}^\dagger}]$. Here a is the grid mesh size and $t_{de}$ is defined as $t_{de} = \frac{\hbar^2}{2m_{de}} a^2$ where $m_{de}$ is the electron effective mass in the dielectric. $V_{eff}$ is the effective voltage at the trap position. When a voltage is applied across a dielectric layer the entire voltage can be assumed to appear and be uniformly distributed across that layer. Hence, the effective voltage at different trap positions is different. Effective voltage has a power factor α over it. The power factor appears because the current due to soft breakdown of dielectric layer exhibits power law dependence on the corresponding voltage. Hence, for soft dielectric breakdown, α is greater than 1. However, when there is a hard breakdown α is equal to 1. Therefore, when a percolation path is formed, the type of breakdown needs to be determined first and the α factor is tuned accordingly. Also the value of α depends on the dielectric material and the capture cross section of the trap. For larger capture cross section α is larger. Here one thing worth mentioning is that this α is different from the voltage acceleration factor appearing in analytical equations for soft breakdown ($I \propto V_g^\gamma$, γ is called the voltage acceleration factor [20]). Therefor they may have different value. In eqn. (5), k is the wave vector. It can be defined as:

$$k = \cos^{-1}(1 - \frac{E - qV_{eff}^{\alpha} + \xi i}{2t_{de}})$$  (6)

$\xi$ is a very small arbitrary energy. One thing worth mentioning is that except hard breakdown we should not add the self-energy matrices of Büttiker probes to the system Green's function. This is because Büttiker probes are artificial probes introduced for modeling. They are not the part of actual physical system. Hence, introduction of Büttiker probes should not disturb the system Green's function. But when there is a hard breakdown, the channel's transport characteristic is changed from tunneling to Ohmic conduction. Hence the system's Green's function is modified according to the following equation

$$G^R = [EI - H - \Sigma_L - \Sigma_R - \Sigma_1 - \Sigma_2 - \cdots - \Sigma_n]^{-1}$$  (7)

Here $\Sigma_1, \Sigma_2, \ldots, \Sigma_n$ are the self-energy matrices of the Büttiker probes attached along the channel.

From figure 4(b), let us consider other trap assisted tunneling paths. Electrons can be captured first by the trap at grid position 2 and then it can tunnel to the right contact. In this case the tunneling transmission coefficient

$$\frac{1}{T_{L2R}} = \frac{1}{T_{L2}} + \frac{1}{T_{2R}}$$  (8)

where $T_{2R} = Trace\ [\Gamma_2 G^R \Gamma_R G^A]$.

Similarly, as stated earlier, electron can travel from left contact to the trap at grid position 5 and then to right contact (fig 4(b)). Hence, the tunneling transmission coefficient

$$\frac{1}{T_{L5R}} = \frac{1}{T_{L5}} + \frac{1}{T_{5R}}$$  (9)

Here $T_{L5} = Trace\ [\Gamma_L G^R \Gamma_5 G^A]$. Note, the electron can tunnel directly from left contact to right contact. Hence, there are a total of 4 paths for electrons to tunnel from left to right contact (fig 4(b)). As all these paths are parallel, the overall tunneling transmission coefficient

$$T_{all} = T_{L25R} + T_{L2R} + T_{L5R} + T_{LR}$$  (10)

The total tunneling current can be expressed as

$$J = \frac{q}{h} \int T_{all}\ (f_R - f_L)dE.$$  (11)

The Büttiker probe is used to model the percolation path (fig 4c). Post breakdown I-V characteristic can be modeled in the same way by attaching Büttiker probe to all the trap locations along the thickness (fig 4d). The carriers have many probable paths for trap assisted tunneling. All the tunneling probabilities are calculated as describe earlier. Then they are added together to get the total transmission probability. In case of a MOS device, the channel region acts as the right contact (fig 4(c)). Therefore, the potential profile in the channel region should be carefully modeled and mapped into the system Hamiltonian. Especially if a drain voltage is applied then the potential profile becomes asymmetric near the source and drain regions. In order to take that into consideration, we can divide the gate into small areas in which the potential profile can be assumed constant.

The current through each of the small areas can be determined individually and integrated to get the total gate tunneling current. In addition, we need to consider the effect of flat band voltage and charge accumulation profile in the channel before and after inversion. If the dielectric layer consists of multiple oxide materials, potential profile in each of the materials needs be determined and included in the Hamiltonian accordingly. Using a similar approach, we can determine the spin current in an MTJ before and after the breakdown. The spin current density between two successive lattice points is defined as [21] [22]

$$J_s = \frac{2\pi}{ih} \int Real[Trace(S. (HG^n - G^n H))]\ dE.$$  (12)

where, H is the system Hamiltonian and S is Pauli matrix. We need to determine the z-oriented spin current if we assume that both the fixed and free layers are pointing in the z direction.

One important thing is that the magnetic contact self-energy contains the exchange splitting energy, $\Delta$. But trap's electron capturing capability is independent of spin. Therefore, Büttiker probe self-energy matrix does not contain $\Delta$.

## III. RESULTS AND DISCUSSION

In the dielectrics, stress induced traps due to high electric field eventually lead to formation of percolation paths. In this section, our proposed model is applied to calculate the stress-induced leakage current (SILC) due to the formation of multiple traps in HfSiON. It is followed by calculation of soft breakdown (SBD) and hard breakdown (HBD) currents in SiO2. The results are in close agreement with the experimental data. Then we consider the formation of percolation paths in MgO. We apply our simulation framework to calculate the post breakdown I-V characteristics and compare with experimental results. We have also calculated the TMR degradation in MTJ due to the formation of percolation paths in MgO. Finally, we have estimated the lifetime of the MTJ based on our calculation. Simulation parameters used in this paper are listed in table I.

### A. Stress Induced Leakage Current (SILC)

Amorphous dielectrics are being used extensively in modern electronic devices. Moreover, these dielectrics are thin and exposed to very high electric field. Therefore, it is very important to analyze the effect of traps and point defects, both in the bulk and on the surface of these dielectrics. Traps can be either pre-existing or they can form over time due to stress (electric field stress or temperature stress). Here we first analyze the traps formed due to stress from high electric field. We will show how to calculate the stress induced trap density using Büttiker probe simulation by matching post-stress $I_g$-$V_g$ characteristics. Then we will discuss the theory of calculating the pre-existing defect density using our simulation framework. We have initially considered high-k metal gate oxide (HKMG) transistor with a defect free 1.8 nm HfSiON layer and 1 nm SiO2 interface layer (IL). First we have matched the pre-stress I-V characteristic with experimental data from [23] (root mean square error < 5%). Simulation parameters are listed in table I. In order to create stress induced defects and observe SILC, a constant voltage stress is applied on this dielectric stack for a very short time [23]. The dielectric layer is quite thick and free of pre-existing defects. Therefore, application of a stress voltage for a very short time will generate only a few traps. The work presented in [23] does not provide any information regarding the number of traps or their locations. The probability of stress-induced trap formation in a dielectric stack is higher at the interface of the two dielectrics [24]. Hence, in this case, we have assumed that the traps will form at the interface of HfSiON and SiO₂ layers. Regarding the energy level of the traps, we have considered the same distribution (fig. 5a,5b) as shown in ref [25] [26].

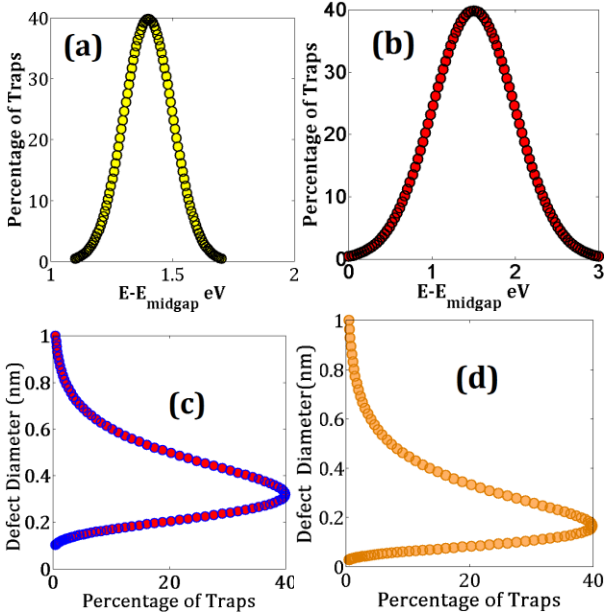

Fig 5. Trap energy distribution of (a) HfSiON (similar to ref [25]) and (b) SiO₂ (similar to [26]). Corresponding defect diameter distribution of (a) HfSiON and (b) SiO₂

The capture cross section dimension of a trap depends on its energy level. The mid bandgap traps of HfSiON and SiO₂ layers have capture cross section diameters of ~ 1nm [12] and 0.6nm [12], respectively. On the other hand, traps on the conduction band have zero capture cross section area since they do not contribute to trap assisted tunneling. In thin oxides, most of the traps have their energy level distributed near the conduction band [25] [26] (fig. 5). Therefore, in thin dielectric layers, the capture cross sections of mid-gap traps to conduction band traps are assumed to decrease exponentially. The exponential function of decreasing capture cross section diameter is modelled as d=exp(-a). Here d is the capture cross section diameter of a trap and a is a fitting parameter which satisfies the previous statement. Depending on the energy level of a trap, the capture cross section distribution can be found from the exponential function (Fig 5c and 5d). One Büttiker probe is placed at each trap positions to calculate the post-stress trap assisted tunneling. The number of Büttiker probes at the HfSiON/SiO2 interface and the acceleration factor α (in eqn. 6) are calibrated for the best match with the experimental data. The parameter, α depends on the capture cross section diameter of the corresponding trap. α is bigger for the trap with larger cross section. For HfSiON, α is calibrated to be between 10 and 11 while for SiO₂, α is distributed between 1.5 to 1.9. We have found that for a trap density of 1.5x10¹⁹/m³ at the HfSiON and SiO2 interface, the calculated SILC matches best with the experimental data (fig. 6). Thus using the Büttiker probe method, we can determine the stress induced trap density.

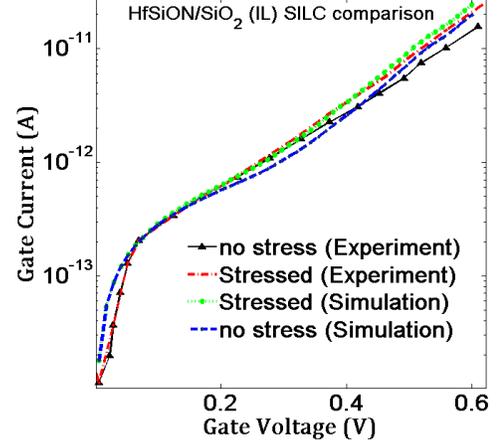

Fig. 6. Pre and post stress induced leakage current for 1.4 nm HfSiON and 0.8nm SiO₂ IL (both experiment [23] and simulation). The root mean square error is less than 5%.

The pre-existing defect density in the oxide layer can be determined in a similar way. Pre-stress defect free I-V characteristic for a different sample of HfSiON/SiO₂ dielectric stack (same or different dimension) can be determined form the same NEGF simulation framework. Let us assume that experimentally observed current is larger than simulation value at all voltages for the new device. This extra current is coming from pre-existing trap assisted tunneling. For introducing this trap assisted tunneling into NEGF simulation framework, we need to include one Büttiker probe at each defect position. The number, positions and self-energies (i.e., value of α) of these probes are calibrated in the simulation framework for matching the experimental I-V characteristics. From the number of Büttiker probe needed in simulation for matching the pre-stress I_V characteristics, we can calculate the pre-existing defect density.

### B. SiO₂ Soft Breakdown

It has been experimentally observed that the SBD in dielectrics is characteristically different from the HBD [6]. In case of soft dielectric breakdown, the post breakdown current shows comparatively smaller increment from the pre-breakdown value and follows power law. Soft breakdown can be modeled with our proposed Büttiker probe by attaching probes to all grid points along the breakdown path (fig. 4d). The shape of the breakdown path (how many traps line up to form the percolation path) can be determined by the percolation theory. The energy levels of the traps are assumed to have Gaussian distribution from conduction band edge to 3 eV inside the bandgap (fig. 5b) [26]. If we tune the power factor α in equation (5), we can get excellent matching of post breakdown I-V characteristics. Here, one thing worth mentioning is that the power factor α depends on the capture cross section of the defects. As mentioned earlier, capture cross section depends on the energy level of the traps. Traps at the percolation path can form at different energy levels. Therefore, α can be different for different Büttiker probes. In ref [16], the post soft breakdown I-V characteristic in a MOS capacitor (Poly-Si/SiO₂/Si system) has been experimentally demonstrated. While benchmarking this simulation, we have observed that most of the traps have the power factor α = 1.7. Few traps have α near 1.5 and these traps can be assumed to be close to the conduction band (fig 5b). These traps have lower capture cross section and contribute less towards tunneling. While few other traps have α near 1.9 and these traps can be assumed to sit a bit deeper (close to mid-bandgap) in energy diagram. These traps have bigger capture cross section and contribute more towards tunneling. Applying these values of α, we have seen excellent match between our simulation result and the experimental observation (shown in fig 7a). Simulation parameters are listed in table I.

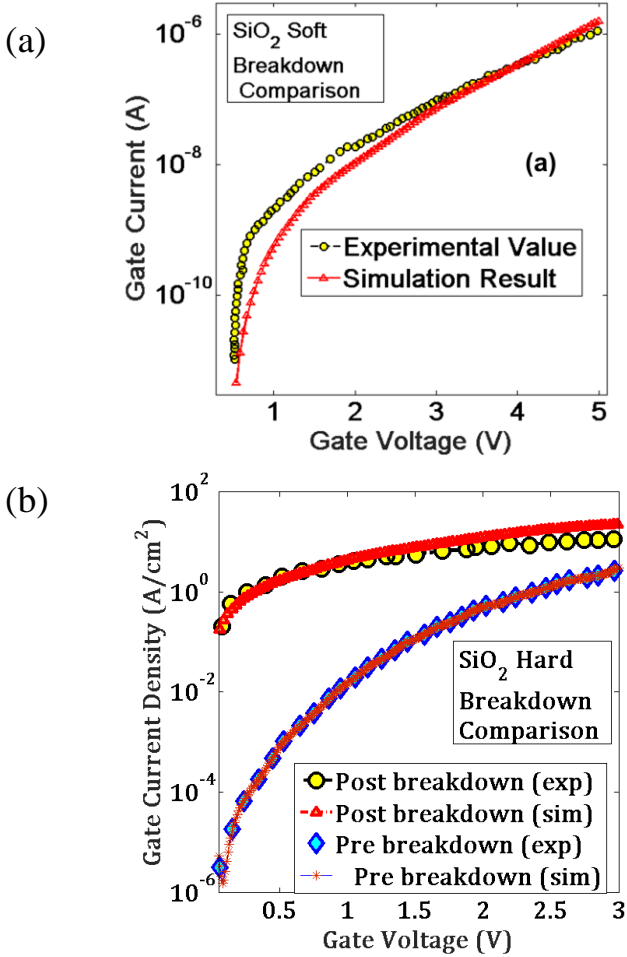

Fig. 7(a). Post soft breakdown current for 1.7 nm SiO₂ (both experiment [16] and simulation). Root mean square of error is 11.76% (b). Post hard breakdown current for 2.2 nm SiO₂ (both experiment [27] and simulation). ). Root mean square of error is ~13%

### C. SiO₂ Hard Breakdown

Hard breakdowns mainly occur in thick dielectrics [6] because of high gate voltage. It can also happen in thin dielectrics if the electric field stress is very high. A hard breakdown can be easily separated from a soft breakdown by the huge change in post-breakdown (PBD) current and the Ohmic nature of the PBD I-V characteristic (the power factor α = 1). In this case, the system's retarded Green's function will be constructed according to eqn. 7. In ref [27], post hard breakdown I-V characteristic is shown for a MOS capacitor with 2.2nm thick SiO₂. In figure 7(b), we have shown both the experimental and simulation data before and after the hard breakdown (simulation parameters are listed in table I). It is evident that after the hard breakdown, the gate leakage current increases by several orders of magnitude but the increase of current with voltage is almost linear.

### D. MgO Soft Breakdown

MgO is a mid-κ dielectric with a low Weibull slope [28]. It has gained popularity due to its excellent spin current filtering capacity. Hence it has become the prime choice for fabricating MTJ. In [29], a post soft breakdown I-V characteristic of 1nm thick MgO in an MTJ is reported. The MTJ stack is comprised of CoFeB/MgO/CoFeB. It was observed in [30] that the SILC activation energy of MgO (0.37eV) is quite high (compared with HfO₂ [24]). Therefore, although the bulk trap has large capture cross section (~ 0.9 nm [28]), in a very thin MgO layer, the trap energy levels are distributed near conduction band edge. As a result, the trap capturing cross section will be much smaller than the bulk value like other dielectrics. Considering this fact, we have calibrated the post-soft breakdown I-V characteristics with the experiment [29]. Most of the traps have the power factor α to be equal to 9. Traps energy level distribution is again assumed to be Gaussian, distributed from the bottom of conduction band to 2.5eV inside the band gap. Most of traps are assumed to be at an energy level of 1.25 eV. Very few traps near the conduction band have α to be around 8.5. Few traps deep inside the bandgap have α to be around 9.5.

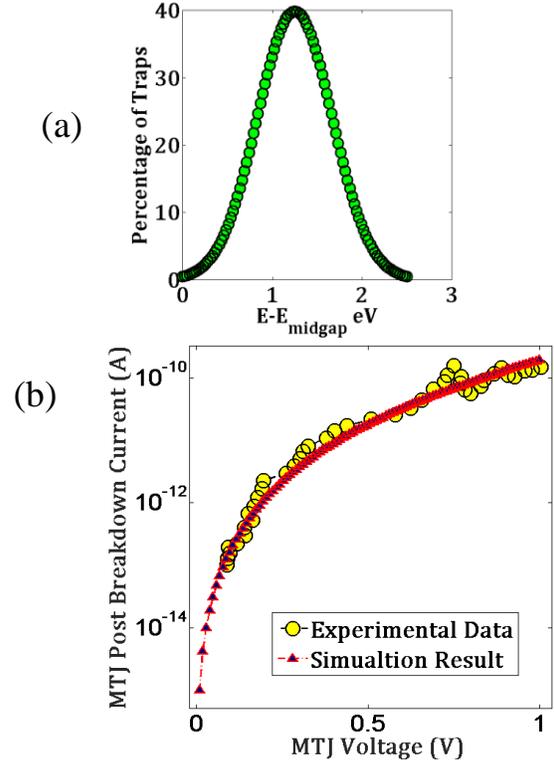

Fig. 8 (a) Trap energy distribution of MgO (b)Post soft breakdown current voltage characteristics in MTJ from experiment [29] and from our simulation framework (root mean square error is less than 2%). Breakdown in MgO layer has a significant impact on the MTJ characteristics. Due to breakdown, the %TMR decreases significantly. This is because the parallel tunneling magnetoresistance (R_P) is mainly dominated by the exchange splitting energy, Δ of the magnetic contacts (shown in fig. 1). For parallel configuration, huge exchange splitting allows the majority of spin to tunnel easily through the energy barrier created by MgO layer. The formation of a percolation path does not affect this process significantly. Hence, $R_P$ does not changes significantly due to breakdown. But for the antiparallel configuration this exchange splitting energy opposes spin tunneling. Therefore, a percolation path significantly decreases the antiparallel tunneling magnetoresistance ($R_{AP}$). In figure 9(a) we have shown the pre and post-breakdown parallel and antiparallel tunneling magnetoresistance ($R_P$ and $R_{AP}$ respectively) of a CoFeB-MgO-CoFeB MTJ with 1nm thick MgO layer (same structure as shown in [3]). Also the corresponding degradation in TMR is shown in figure 9(b). We can observe that $R_{AP}$ decreases at a much faster rate than $R_P$.

TABLE I
PARAMETERS USED IN BÜTTIKER PROBE SIMULATION

| Parameter Name | Value |
|---|---|
| Grid mesh size | 0.1 nm |
| CoFeB effective mass ($m_{FM}/m_0$) | 0.8 [32] |
| CoFeB-MgO barrier height | 0.77 eV [32] |
| CoFeB-MgO Fermi level | 2.25 eV [32] |
| CoFeB exchange splitting energy | 2.15 eV [32] |
| MgO tunneling effective mass ($m_{FM}/m_0$) | 0.18 [32] |
| Weibull slope of MgO ($\beta$) | 0.6 [30] |
| MgO voltage acceleration factor ($\gamma$) | 25.6 [30] |
| TiN work function | 4.4 eV [33] |
| TiN effective mass ($m_{TiN}/m_0$) | 1.1 [34] |
| Si effective mass | 0.26 |
| Si electron affinity | 4.05 eV |
| Temperature | 300 k |
| HfSiON tunneling effective mass | 0.03 [35] |
| HfSiON bandgap | 5.32 eV [35] |
| HfSiON electron affinity | 2.95 eV [35] |
| TiN-HfSiON barrier height | 1.45 eV |
| SiO₂ tunneling effective mass | 0.42 [36] |
| SiO₂ electron affinity | 0.95 eV [37] |
| Si-SiO₂ barrier height | 3.1 eV [38] |
| SiO₂ bandgap | 9 eV |

The reason of %TMR degradation after breakdown can be attributed to the spin independent nature of the dielectric traps [4][5]. Traps inside the MgO layer helps tunneling irrespective of electron spin. When an electron is captured by a trap its spin properties get randomized. The spin wave function of an electron in a trap can be represented as

$$|\psi_s\rangle = a_{z+}|u_{z+}\rangle + a_{z-}|d_{z+}\rangle \qquad (13)$$

Here $a_{z+}a_{z+}^*$, $a_{z-}a_{z-}^*$, $|u_{z+}\rangle$ and $|d_{z+}\rangle$ represents the spin polarization probability and spin basis vector (spinors) along +z and -z axis, respectively. These two spinors can represent spin pointing at any direction. In a trap it can be assumed that these two probabilities are equal i.e., $a_{z+}a_{z+}^* = a_{z-}a_{z-}^*$. Therefore, the spin filtering efficiency of the MgO layer is expected to go down after the formation of traps and percolation paths. Fig. 9 (b) shows the TMR degradation of an MTJ operating at different voltages. If the MTJ operates at 1mA (V~0.6V), we observe that the TMR goes down by about 25% after one soft breakdown. Therefore, 3 to 4 soft breakdowns in the MgO layer can cause functional failure of the MTJ.

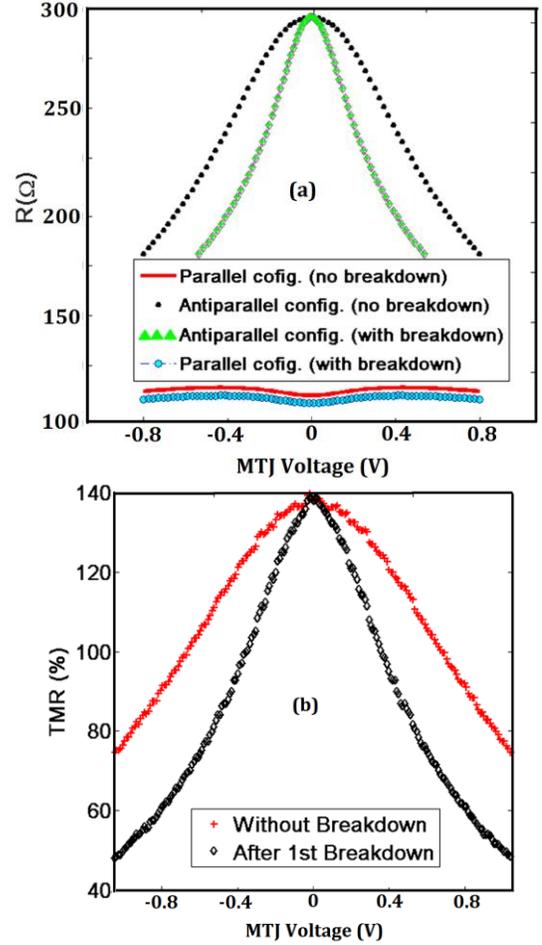

Fig.9(a): Simulation data for pre and post breakdown (bd) parallel and antiparallel resistance ($R_P$ and $R_{AP}$) in MTJ (device dimension can be found in [3]) (b) Change in %TMR before and after breakdown (simulation data)

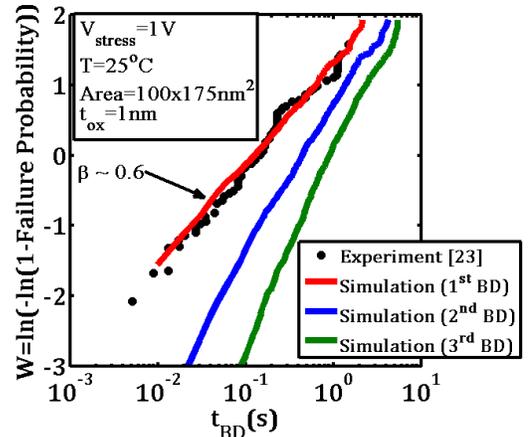

Fig.10: Weibull plot of MgO for 1st and 3rd breakdown. 1st breakdown data is calibrated with experiment [30] using percolation model and based on the calibrated data, Weibull plot for 3rd breakdown is plotted at a stress voltage of 1V.

In order to estimate the lifetime of MTJs, we have implemented a 3-D cell based percolation model [12] and calibrated the Weibull distribution after the 1st breakdown with experimental data [30]. The simulation is done using 1000 sample MTJs. Then the Weibull plot for 3rd SBD is drawn (fig.

10) using data from the percolation model. We can see that 1% of the MTJs suffer $3^{rd}$ SBD after 0.03s for a stress voltage of 1V. As the MgO area in [30] is different from our simulation, we need to do area scaling using standard area scaling formulation [8]. After area scaling, we have found that 1% of the 1000 MTJ samples have $3^{rd}$ SBD after 0.0631s at stress voltage of 1V. For determining the device lifetime at operating voltage, we need to do voltage scaling. For voltage scaling we have used the following equation [8],

$$V_{op} = V_{acc} - log_{10}(\frac{t_{op}}{t_{acc}})/\gamma \qquad (14)$$

Here, $\gamma$ is the voltage acceleration factor (value is listed in table I). We can see that at operating voltage of 0.6V, 1% MTJs in 1000 sample MTJs will have $3^{rd}$ soft breakdown in almost 24 years. Therefore, we can draw the conclusion that MgO based MTJs have comparatively much longer lifetime than standard CMOS devices.

## IV. CONCLUSION

In this work, we have presented a Büttiker probe based post breakdown current model for dielectric materials. The proposed method is flexible, can be applied to a wide range of dielectric materials, and has shown excellent potential for TDDB analysis. The simulation framework shows very good match with experimental data for any type, shape and size of dielectric. In addition, it has the flexibility to model traps formed at any position. It is a physics based, close to atomistic simulation model, yet does not consume much computational resources. Most importantly, it can predict spin current and TMR degradation in MTJs that cannot be done with conventional TDDB models.

## REFERENCES

[1] Büttiker, M. "Four-Terminal Phase-Coherent Conductance," Phys. Rev. Lett., 57:1761–1764, 1986..

[2] S. Datta, *Quantum Transport*, 2nd ed. Cambridge, U.K.: Cambridge Univ. Press, 2005.

[3] H. Kubota, et al, "Quantitative measurement of voltage dependence of spin-transfer torque in MgO-based magnetic tunnel junctions", Nat. Phys., vol. 4, no. 1, pp. 37-41, 2008.

[4] Chevy F, Madison K, Bretin V and Dalibard J. *Trapped Particles and Fundamental Physics*. Berlin: Springer, 2002, pp 92

[5] Zongli Sun and QiangGu. "Spontaneous separation of large-spin Fermi gas in the harmonic trap: a density functional study", Scientific Reports **6**, Article number: 31776, 2016

[6] M. A. Alam, B. E. Weir, and P. J. Silverman, "A study of soft and hard breakdown -,Part I: Analysis of statistical percolation conductance", IEEE Trans. Electron Devices, vol. 49, pp. 232-238, 2002

[7] B. Weir, C. Leung, P. Silverman and M. A. Alam, "Gate dielectric breakdown in the time-scale of ESD events," Microelectron. Reliab. vol.45, pp.427-436, 2005.

[8] M. A. Alam, B. E. Weir, and P. J. Silverman, "A study of soft and hard breakdown -,Part II: Principles of area, thickness, and voltage scaling", IEEE Trans. Electron Devices, vol. 49, pp. 239-246, 2002

[9] M. A. Alam, et al, "Explanation of soft and hard breakdown and its consequences for area scaling", IEDM Tech. Digest, pp. 449-452, 1996.

[10] M. A. Alam, "SILC as a Measure of Trap Generation and Predictor of TBD in Ultrathin Oxides", IEEE Transaction on Electron Devices, 49 (2), pp. 226-231, 2002.

[11] M. A. Alam, B. E. Weir, and P. J. Silverman, "A Study of Soft and Hard Breakdown : The Statistical Model," IEEE Transaction on Electron Devices, 49 (2), pp. 232-238, 2002.

[12] T. Nigam, A. Kerber, and P. Peumans "Accurate model for time-dependent dielectric breakdown of high-κ metal gate stacks," in Proc. IEEE IRPS, 2009, pp. 523-530.

[13] M. Masuduzzaman, A.E. Islam and M.A. Alam, "Exploring the capability of multi-frequency charge pumping in resolving location and energy levels of traps with dielectric", IEEE Trans Electron Dev, 55 (12) (2008), pp. 3421–3431

[14] K. Okada, et al, "A new prediction method for oxide lifetime and its application to study dielectric breakdown mechanism," in *Proc. Dig. Symp. VLSI Technol.*, 1998, pp. 158–159.

[15] T. Nigam, S. Martin, and D. Abusch-Magder, "Temperature dependence and conductionmechanism after analog soft breakdown," in *Proc. Int. Rel. Phys. Symp.*, 2003, pp. 417–423.

[16] A. Cester, et al, "A novel approach to quantum point contact for post soft breakdown conduction," in *Proc. IEDM Tech. Dig.*, 2001, pp. 305–308.

[17] R. P. Vedula, S. Palit, M. A. Alam, and A. Strachan, "Role of Atomic Variability in Dielectric Charging: A First Principles-based Multi-scale Modeling Study",Physical Review B, vol. 88 (20), 2013, pp. 205204.

[18] M. Julliere (1975). "Tunneling between ferromagnetic films". Phys. Lett. 54A: 225–226.

[19] Datta, S. *Lessons from Nanoelectronics*: A New Perspective on Transport (World Scientific, 2012).

[20] L. A. Escobar and W. Q. Meeker, "A review of accelerated test models," *Statistical Science*, vol. 21, no. 4, pp. 552–577, 2006.

[21] D. Datta , B. Behin-Aein , S. Salahuddin and S. Datta, "Quantitative model for TMR and spin-transfer torque in MTJ devices ", Proc. IEEE Int. Electron Device Meeting, pp. 22.8.1-22.8.4, Dec. 2010.

[22] A. K. Reza et al, "Modeling and Evaluation of Topological Insulator/ Ferromagnet Heterostructure-Based Memory", IEEE TRANSACTIONS ON ELECTRONS DEVICES, VOL. 63, NO. 3, MARCH 2016

[23] R. O'Connor , L. Pantisano and R. Degraeve, "SILC defect generation spectroscopy in HfSiON using constant voltage stress and substrate hot electron injection", *Proc. IEEE IRPS*, pp. 324-329, 2008.

[24] K. S. Yew, D. S. Ang, and G. Bersuker, "Bimodal Weibull distribution of metal/high-K gate stack TDDB—Insights by scanning tunneling microscopy," IEEE Electron. Device Lett., vol. 33, no. 2, pp. 146–148, Feb. 2012

[25] M. Duan, et al, "Insight Into Electron Traps and Their Energy Distribution Under Positive Bias Temperature Stress and Hot Carrier Aging", IEEE TRANSACTIONS ON ELECTRON DEVICES, VOL. 63, NO. 9, SEPTEMBER 2016

[26] Nathan L. Anderson, et al, "Defect level distributions and atomic relaxations induced by charge trapping in amorphous silica", Applied Physics Letters 100, 172908 (2012).

[27] Kenji Komiya and Yasuhisa Omura, "Aspects of Hard Breakdown Characteristics in a 2.2-nm-thick SiO2-Film", Journal of Semiconductor technology and science, vol. 2, No. 3, September, 2002

[28] R. O'Connor, G. Hughes, and P. Casey,"Degradation and breakdown characteristics of thin MgO dielectric layers" J. Appl. Phys. 107, 024501, 2010.

[29] C. Yoshida, et al, "A study of dielectric breakdown mechanism in CoFeB/MgO/CoFeB magnetic tunnel junction," in Proc. Int. Rel. Phys. Symp. (IRPS), Montreal, QC, Canada, 2009, pp. 139–142.

[30] C. Yoshida and T. Sugii, "Reliability study of magnetic tunnel junction with naturally oxidized MgO barrier," in Proc. IEEE IRPS, Apr. 2012, pp. 2A.3.1–2A.3.5..

[31] J. Yang, et al, "Intrinsic correlation between PBTI and TDDB degradations in nMOS HK/MG dielectrics," in Reliability Physics Symposium (IRPS), 2012 IEEE International, April 2012, pp. 5D.4.1 – 5D.4.7

[32] D. Datta, et al, "Voltage asymmetry of spin-transfer torques," IEEE Trans. Nanotechnol., vol. 11, no. 2, pp. 261–272, Sep. 2012.

[33] L. P. B. Lima, et al, "Metal gate work function tuning by Al incorporation in TiN," J. Appl. Phys., vol. 115, no. 7, p. 074504, 2014

[34] J. S. Chawla, X. Y. Zhang, and D. Gall, "Effective electron mean free path in TiN(001)," J. Appl. Phys. 113(6), 063704 (2013)..

[35] M.J. Chen and Chih-Yu Hsu, " Evidence for a very small tunneling effective mass (0.03 m₀) in MOSFET high-k (HfSiON) gate dielectrics ", EDL., Vol. 33, No. 4, p. 468, 2012.

[36] B. Brar, G. D. Wilk, and A. C. Seabaugh, "Direct extraction of the electron tunneling effective mass in ultrathin SiO2," Appl. Phys. Lett., vol. 69, no. 18, pp. 2728–2730, Oct. 1996.

[37] Y.-C. Yeo, T.-J. King, and C. Hu, "metal-dielectric band alignment and its implications for metal gate complementary metal–oxide–semiconductor technology," J. Appl. Phys., vol. 92, pp. 7266–7271, 2002.

[38] A. Hadjadj, et al., "Si–SiO2 barrier height and its temperature dependence in metal-oxide-semiconductor structures with ultrathin gate oxide," Appl. Phys. Lett., 80(18), 3334-3336, 2002.